\documentclass{article}

\usepackage{amsmath, amssymb} 
\usepackage[UglyObsolete,tight,heads=LaTeX] {diagrams}
\newcommand{\bc}{\begin{C}}
\newcommand{\ec}{\end{C}}
\newcommand{\be}{\begin{equation}}
\newcommand{\ee}{\end{equation}}

\newcommand{\claim}{\begin{Cl}}
\newcommand{\eclaim}{\end{Cl}}
\newcommand{\nb}{\begin{Nb}}
\newcommand{\nbe}{\end{Nb}}
\newcommand{\bl}{\begin{LE}}
\newcommand{\el}{\end{LE}}

\newtheorem{Cl}{Claim}

\newcommand{\bd}{\begin{Def}}
\newcommand{\ed}{\end{Def}}
\newcommand{\bt}{\begin{Th}}
\newcommand{\et}{\end{Th}}

\newtheorem{Th}{Theorem}
\newtheorem{LE}{Lemma}
\newtheorem{C}{Corollary}
\newtheorem{Nb}{Note}
\newtheorem{Def}{Definition}

\def\Id{\rm id}

\begin{document}
 \title{On Banach spaces of sequences and free linear logic exponential modality
}
\author{Sergey Slavnov
\\  National Research University Higher School of Economics
\\ sslavnov@yandex.ru\\} \maketitle

\begin{abstract}
We introduce a category of vector spaces modelling full propositional linear logic, similar to probabilistic coherence spaces and to Koethe sequences spaces. Its objects are {\it rigged sequence spaces}, Banach spaces of sequences, with norms defined from pairing with finite sequences, and morphisms are bounded linear maps, continuous in a suitable topology. The main interest of the work is that our model gives a realization of the free linear logic exponentials construction.

\end{abstract}

\section{Introduction}
It is customary to interpret linear logic metaphorically in the language of vector spaces; such an intuition of course is implied in the very title ``linear''. For separate fragments we can do better than metaphors and build from this intuition concrete mathematical models fairly easily. However, it has been noted for a while that the full linear logic with all its three layers of connectives (multiplicative/additive/exponential) somewhat resists consistent interpretation in this context: the layers seem hard to reconcile.

Indeed, the multiplicative layer with tensor product, linear implication and linear negation is intuitively interpreted as corresponding to, respectively, tensor product of vector spaces, vector space of linear maps and vector space duality; the latter should be involutive, which may be problematic in infinite dimensions. On the other hand, the additive layer is  understood as corresponding to normed spaces with the product represented by $l^\infty$-norm and coproduct, by $l^1$-norm. Finally, the exponentials require some  ``Fock space'' of power series, i.e. some completion of the free symmetric algebra (as discussed in \cite{BlutePanangadedOldFoundations}, see also \cite{Girard99coherentbanach}). It is not clear how and under what conditions we can equip the space of power series with a norm satisfying necessary properties.

A systematic attempt was done in Girard's work on {\it coherent Banach spaces} \cite{Girard99coherentbanach}, where a pair of dual formulas  $A,A^\bot$ is interpreted as a {\it dual pair}  $(A_+,A_-)$ of Banach spaces equipped with a non-degenerate pairing, such that norm on each of the two spaces is  determined from pairing with the other one.  Negation, or duality, corresponds then to exchanging $A_+$ and $A_-$. It is necessary to choose simultaneously a pair of Banach spaces for a formula and for its dual in order to  ensure involutivity of negation; recall that, in infinite dimension, the usual Banach space duality, in general, is not involutive.

This allows fairly reasonable interpretation of multiplicative-additive connectives in the style, discussed in the beginning. In order to model exponentials it was necessary to consider, the space of {\it analytic} functions,  (i.e., functions defined by converging power series). A natural norm for such a space would be  the sup-norm, however non-constant functions, analytic on an unbounded domain, are necessary unbounded, therefore Girard considered functions analytic in  the {\it open} unit ball. Unfortunately, functions arising from linear logic proofs tend to map the interior of the unit ball to its closure, hence composing them was questionable (and there is no reasonable way to extend a general analytic function from the open ball to its boundary). The solution proposed in \cite{Girard99coherentbanach}  was  to drastically and in a somewhat {\it ad hoc} way modify the {\it logic} in order to fit the model, but interpretation of  the usual linear logic definitely failed.

Probably the most known and successful interpretation of full linear logic in the setting of vector spaces is the {\it Koethe sequences} model developed in \cite{Ehrhard_Koethe}. (The current work is to a large extent  inspired by \cite{Ehrhard_Koethe} and borrows many ideas from there.)  However, in that model, the problem with additives is simply ignored: Koethe sequence spaces do not have norms at all, but, also, they do not have any distinction between products and coproducts. In other words, the dual additive connectives are identified, and the model is degenerate. Which is not completely satisfactory.
A similar degeneracy occurs in the category of {\it convenient vector spaces} \cite{BluteEhrhardTasson}, which models {\it differential linear logic} (see \cite{diff-lambda}, \cite{diff-nets}), where exponential connectives play a crucial role.

A vector space-like model that supports full linear logic, including nondegenerate additives, is the category of {\it Probabilistic coherence spaces} (PCS), initiated in \cite{Girard_Logic_quantic} and developed in \cite{DanosErhahrdPCS}. However, PCS are not precisely vector spaces (although any PCS gives rise to a Banach space, as described in \cite{DanosErhahrdPCS}).  In particular, PCS morphisms  in \cite{DanosErhahrdPCS} do not have  any clear description in  the intrinsic language of Banach spaces, moreover, it is shown that there are Banach spaces morphisms that do not correspond to any PCS morphism.  Thus, a relation between PCS and corresponding Banach spaces should be clarified.

 In this work we introduce the category of {\it rigged sequence spaces} (RSS), which turns out to be essentially equivalent to PCS. In particular, it supports full propositional linear logic with non-degenerate additives. Our presentation, however, is different, and RSS are defined directly as specific Banach spaces. In particular, RSS morphisms are defined in an intrinsic, coordinate-free way. The main novelty is that we show that exponentials in RSS are {\it free}. This seems important because, unlike the case of multiplicative-additive fragment,  interpretation of linear logic exponential connectives, in general, is not unique. That is, in one model of linear logic, several non-isomorphic exponential modalities may coexist (see \cite{Mellies_categorical_semantics} for a discussion). Thus, free exponentials, if they exist, are ``minimal'' and have very canonical flavour.

 Also, in  \cite{Mellies_free_exp} Melli\'es et al. discovered an abstract construction of  free exponential modality, which applies when the modeling category satisfies certain properties. It is interesting to find concrete realizations of the abstract construction, especially in the setting of Banach spaces, where interpretation of exponentials is not easy. And indeed, we find that the free exponentials in RSS arise  precisely along the lines of \cite{Mellies_free_exp}.

\subsection{Coherence spaces and vector spaces}
The first and best known model of linear logic, coherence spaces, has free exponentials. But the coherence spaces structure can itself be described as a construction on ``vector spaces'' (i.e. free modules) over ${\bf N}$. Indeed, for a coherence space $A$ consider the free ${\bf N}$-module $S_A$ on its web $|A|$, i.e. the space of natural number sequences indexed by points of  $|A|$. All subsets of $|A|$ are represented as elements of $S_A$ by their characteristic functions. The coherence space structure can be recovered from the set of finite cliques (even of two-element cliques) or the set of finite anti-cliques.  Let the obvious semi-finite pairing of elements of $S_A$ be given by $\langle a,b\rangle =\sum\limits_{i}a_ib_i$ (the result is a natural number or $\infty$); define the semi-finite norm on $S_A$ by
$$||a||=\sup\{\langle a,b\rangle |\mbox{ }b\in S_A\mbox{ represents a finite anti-clique}\}. $$
Then the ``unit ball'' $\{a\in S_A|\mbox{ }||a||\leq 1\}$ is exactly the set of (elements representing) cliques. Furthermore, anti-cliques, i.e. cliques of the dual coherence space, correspond precisely to elements of the ``dual ball'': $b$ corresponds  to an anti-clique iff $||b||^*\leq 1$, where
$||b||^*=\sup\limits_{||a||\leq 1}\langle b,a\rangle $.
An important observation is that when computing  supremum in the above formula we can restrict to finitely supported $a$'s, which means that the dual coherence space is defined exactly in the same fashion.

Thus, coherence space can literally be described as a normed ``vector space'' over ${\bf N}$, and cliques, as vectors of norm one. Morphisms,  in this language, become then ${\bf N}$-linear maps, sending unit ball to unit ball (i.e. {\it bounded} with norm less than or equal to $1$) and preserving arbitrary suprema (with $S_A$ seen as a poset under the pointwise ordering). The last condition can be formulated as continuity in a suitable topology.

In this work we mimic the above description replacing spaces over ${\bf N}$ with spaces over ${\bf R}$ (we may consider the complex ground field as well, but it does not add anything interesting). Consider the index set $I$ and two sequences spaces: $S_I$, the space of finitely supported sequences indexed by $I$, and $S_I^*$, the space of all sequences indexed by $I$ (the latter space is the algebraic dual of the former,  which explains the notation). The role of coherence structure is played by a norm $||.||$ on the space $S_I$ of finite sequences. The norm induces the (semi-finite) dual norm $||.||^*$ (which can be infinite for some vectors) on $S_I^*$ and then the initial norm extends to the semi-finite norm on the whole $S_I^*$ by $||a||=||a||^{**}$. The space $A$ of finite norm vectors is a Banach sequences space, whose unit ball plays the role of the set of cliques. For the want of a better term we call this structure a {\it rigged sequence space} (because we get a ``sandwich'' $S_I\subseteq A\subseteq S_I^*$, similar to the case of rigged Hilbert spaces, see \cite{GelfandVilenkin}).

Following the above description of coherence spaces  we define morphisms as bounded linear operators of norm less than or equal to $1$ and, in addition, continuous in a suitable topology (not the norm-topology). More precisely, our morphisms have decompositions in {\it positive} and {\it negative} parts, $M=M^+-M^-$, which allows defining the {\it absolute value} of the morphism $|M|=|M^+|+|M^-|$, and we require  the absolute values of morphisms to be bounded. The notion of positivity plays an important role in our category, especially, when exponentials and power series  come into play.

We choose specific topology on our spaces so  that continuous maps correspond to matrices. Then every hom-set gets itself the structure of a rigged sequence space. This gives us a $*$-autonomous category with products, hence a model of multiplicative-additive linear logic.

It turns out that, for the category of rigged sequences spaces, which looks so similar to coherence spaces, we manage to mimic the free exponentials construction as well. The exponential spaces are certain subspaces of the direct products of symmetric tensor powers of the original spaces; seen as sequences, their elements are indexed by finite multisets of the original index set. In particular, elements of $?A$ have the expected interpretation as  power series (i.e. real analytic functions) on the dual $A^*$ of $A$, absolutely converging everywhere in the unit ball (including the boundary). This analytic functions interpretation is not extremely exciting, because at the boundary they, in general, become discontinuous, which is also expected. An essential point that makes the construction work is that these functions  have decompositions into positive and negative parts. This allows defining a norm so that elements of $?A$ are dominated by positive power series with uniformly bounded sums on the whole closed unit ball; thanks to this, they have canonical extensions from the open ball to the boundary (possibly discontinuous). This is a way to deal with problems encountered in the seminal work \cite{Girard99coherentbanach} on coherent Banach spaces.

\subsection{Notation and background}
We assume that the reader is familiar with linear logic (see \cite{Girard}, \cite{Girard2} for an introduction), as well as with its interpretation as a $*$-autonomous category (\cite{Seely} is the standard reference, see also \cite{Mellies_categorical_semantics}).

We denote the monoidal product in a $*$-autonomous category as $\otimes$ and call it {\it tensor} product, and we denote duality as a star $(.)^{*}$. The dual of tensor product is called {\it cotensor} and denoted as $\wp$, i.e. $A\wp B=(A^*\otimes B^*)^*$. Monoidal unit is denoted ${\bf 1}$, with its dual ${\bf 1}^*$ denoted $\bot$. Product and coproduct are denoted as, respectively, $\times$ and $\oplus$. The internal homs functor is denoted $\multimap$, as usual.

We also assume that the reader is familiar with basic notions of locally convex vector spaces, in particular, normed and Banach spaces; see any standard textbook, for example \cite{Rudin}.

\section{Rigged sequence spaces}
\subsection{Definition and duality}
Consider an at most countable set of indices $I$. We denote  the space of sequences $a=\{a_i|i\in I\}$, such that only finitely many terms  $a_i$ are non-zero, as $S_I$, and  the  space of all sequences indexed by $I$, as $S_I^*$ ($S_I^*$ is the algebraic  dual of $S_I$). For brevity, we will call  sequences from $S_I$ {\it finitely supported} or, simply, {\it finite} in the sequel.

We have the pairing of sequences given by
\be\label{pairing}
\langle a,b\rangle =\sum\limits_{i\in I}a_ib_i,
\ee
defined for $a\in S_I$, $b\in S_I^*$. This extends to a partial pairing of all sequences, defined if the series in (\ref{pairing}) is absolutely convergent and  undefined otherwise. For non-negative sequences the pairing is always defined as an extended real number from ${\bf R}\cup\{\infty\}$, since a series with nonnegative terms always converges (maybe to infinity). Since for an absolutely convergent series the order of terms does not matter, when computing pairing of sequences in  this Section we will identify the index set $I$ with natural numbers (or a subset of them if it is finite) and omit it from notation.

The space $S_I^*$  (and hence, $S_I$) is partially ordered. A sequence $a$ is {\it nonnegative} if $a_i\geq 0 \forall i\in I$, and $a\geq b$ if $a-b\geq 0$. Furthermore it is in an obvious way a lattice with the maximum $\max(a,b)$ of two sequences $a$ and $b$ being the sequence $\max(a,b)=\{\max(a_i,b_i)\}_{i\in I}$. Thus for any $a\in S_I^*$ there is the {\it positive} and the {\it negative
} part, respectively $a^+=\max(a,0)$, $a^-=\max(-a,0)$. The  operation of {\it absolute value} is defined then by: $|a|=a^++a^-=\{|a_i|\}_{i\in I}$, $|a|\geq 0\forall a$. The operation preserves finite sequences.

For a reader, familiar with coherence spaces: the dual pair $S\subseteq S^*$ plays the role of the {\it web} of a coherence space. Now the role of a coherence relation will be played by a norm on $S$.
We call a  norm on any of the spaces $S$, $S^*$ {\it consistent}, if it satisfies  the following:
\be\label{consistent_norm}
 |b|\geq |a|\mbox{ implies } ||b||\geq||a||.
\ee
(Note that for such a norm we have $  ||a||=||(|a|)||$. )

A {\it rigged sequence space $A$} is defined by a consistent norm $||\mbox{ }||_A$ on $S$.

The consistent norm $|| ||_A$ gives rise to the
 {\it dual norm} on $S$, defined by
\be\label{dual_norm}
||s||_A^*=\sup\limits_{a\in S, ||a||\leq 1}|\langle a,s\rangle |,
\ee
which is easily seen to be consistent as well.
Then the original norm $||\mbox{ }||_A$ has a partial extension to $S^*$, given by
\be\label{norm_ext}
||a||_A=\sup\limits_{s\in S, ||s||^*\leq 1}|\langle a,s\rangle |.
\ee

\bd In notation as above, the rigged sequences space $A$ is the space of sequences $a$ from  $S^*$ for which $||a||_A<\infty$, where $||a||_A$ is defined in (\ref{norm_ext}).

The {\it dual} rigged sequence space $A^*$ of $A$ is defined by the norm $||\mbox{ }||_{A^*}=||\mbox{ }||_{A}^*$, given in (\ref{dual_norm}).
\ed

The following note is very easy.
\nb\label{sup_over_finite_norms}
A Banach space $(A,||\mbox{ }||)$ of sequences on the index set $I$ is a rigged sequence space iff all finite sequences are in $A$, the norm $||\mbox{ } ||$ is consistent, and for any sequence $a$ it holds that
$||a||=\sup\limits_{a'\in S_I,\substack a'\leq A}||a'||$. $\Box$
\nbe
\bigskip

In the sequel we will sometimes denote a rigged sequence space (RSS) $A$ as the triple $S_I\subseteq A\subseteq S_I^*$; however the index set $I$ will be sometimes omitted from notation, when it is clear from context or not important. We also will usually omit the subscript from notation for the norm.

A rigged sequence space is explicitly defined as the dual of a normed space ($A$ is the dual of $(S,||\mbox{ }||_{A^*}$). This immediately yields the following
\nb A rigged sequence space is a Banach space.
\nbe
\bigskip

We now show that  pairing (\ref{pairing}) of elements of $A$ and $A^*$ is well-defined.
\bt
In notation as above, for $a\in A$, $b\in A^*$  the pairing $\langle a,b\rangle $ is a well-defined finite number with the estimate $|\langle a,b\rangle| \leq||a||_A\cdot||b||_{A^*}$.
\et
{\bf Proof.} \hspace{0.01cm} For any sequence $s\in S^*$ let $s^{\leq i}$ be the sequence $\{s_0,s_1,\ldots,s_i,0,\ldots,0,\ldots\}$. That is the sequence, coinciding with $s$ up to the $i$th term, and zero at all other terms.

Without loss of generality we may assume $||a||=||b||=1$. Consider at first the case when $a$, $b$ are non-negative. Then for any $i$ we have $a\geq a^{\leq i}\geq 0$, $b\geq b^{\leq i}\geq 0$, so $||a^{\leq i}||,||b^{\leq i}||\leq 1$, since the norms are consistent. It follows that $\langle a^{\leq i},b^{\leq i}\rangle \leq 1$ for all $i$. But $\langle
a,b\rangle $ is by definition (\ref{pairing}) the limit of the sequence $\{\langle a^{\leq i},b^{\leq i}\rangle \}$. In the case of non-negative $a,b$ this sequence is monotone and, as we have just shown, bounded by $1$, So the limit exists and is bounded by $1$.

For general $a$, $b$, we know already that the pairing $\langle |a|,|b|\rangle $ is well-defined. Which means that the series in (\ref{pairing}) is absolutely converging, so $\langle a,b\rangle $ is well-defined as well, and we have $|\langle a,b\rangle |\leq \langle |a|,|b|\rangle $. Since the norms are consistent, $||a||=||~|a|~||$, $||b||=||~|b|~||$, and the statement follows. $\Box$

\bc For $a\in A$,  it holds that
$||a||_A=\sup\limits_{ b\in A^*, ||b||_{A^*}\leq 1}|\langle a,b\rangle |$.

For $b\in A^*$,  it holds that
$||b||_{A^*}=\sup\limits_{ a\in A, ||a||_A\leq 1}|\langle a,b\rangle |$.
\ec

{\bf Remark}\hspace{0.1 cm}
A rigged sequence space $S_I\subseteq A\subseteq S_I^*$ can be equivalently defined just by specifying the unit balls $B_A=\{a\in S_I^*|~||a||_A\leq 1\}$ in $A$ and $B_{A^*}=\{b\in S_I^*|~||b||_{A^*}\leq 1\}$ in $A^*$, which should satisfy the property
$$B_A=\{a\in S_I^*|~|\langle a,b\rangle|\leq 1 \forall b\in B_A^*\},\mbox{ }B_{A^*}=\{b\in S_I^*|~|\langle a,b\rangle|\leq 1 \forall a\in B_A\}.$$
The reader familiar with probabilistic coherence spaces (PCS) will easily observe that a rigged sequence space is precisely the {\it Banach space associated to a probabilistic coherence space}, see \cite{DanosErhahrdPCS}.

\subsection{Morphisms}
We are going to define morphisms between RSS.

Let us discuss first what do we want from them.

We are working with Banach spaces, so it is reasonable that morphisms should be norm-bounded linear maps. Since we want to have a category with duality, we also want these maps to be {\it adjointable}. That is a map $M:A\to B$ should have the {\it adjoint} $M^*:B^*\to A^*$ defined by $$\langle L^*b,a\rangle =\langle b,La\rangle .$$  Next, we need the internal homs-functor, which means that the space $Hom(A,B)$ of morphisms should be an RSS itself. This suggests that our maps should be  matrices. Also we need to be able to define absolute values of these matrices, and these absolute values  should induce again well-defined bounded adjointable maps.

 These properties actually characterize morphisms completely. However, such a characterization, in particular, representation of maps by matrices, depends on the particular choice of a basis. It is desirable to have a more intrinsic, coordinate-free construction. It turns out that our morphisms can  be described also as maps continuous with respect to a specific topology, which we  now introduce.

Let $A$ be an RSS. We will be interested in the following {\it normal} topology, whose basis neighborhoods of zero are the sets
$$U_{b,\epsilon}=\{a|~\langle |a|,b\rangle < \epsilon\}, $$
where $b\in A^*$, $b\geq 0$, and $\epsilon>0$ is a number.

This is indeed a basis of neighborhoods. If $b,c\in A^*$, $b,c\geq 0$ let $d$ be defined by $d_i=\max(b_i,c_i)$. The sequence $d$ is in $A^*$, since $b+c\geq d\geq 0$, so $||d||\leq||b+c||<\infty$. Then, for numbers $\epsilon_1,\epsilon_2>0$ the intersection $U_{b,\epsilon_1}\cap U_{c,\epsilon_2}$ contains the neighborhood $U_{d,\epsilon}$, where $\epsilon=\min(\epsilon_1,\epsilon_2)$, $\epsilon>0$.

Below, in order to avoid confusion we use the term {\it bounded} for a norm-continuous map, whereas the term {\it continuous} always refers to the normal topology. Now we investigate the structures of continuous maps.

Let $S_I\subseteq A\subseteq S_I^*$, $S_J\subseteq B\subseteq S_J^*$ be RSS. For $i\in I$, $j\in J$ we denote as $e_i$ the sequence from $S_I$ with zeroes everywhere except the $i$th position, where it equals $1$, and as $r_j$, the analogous sequence in $S_J$. For a linear map $M:S_I\to S_J^*$ we denote $M_{ij}=\langle Me_i,r_j\rangle $ and define $\{M_{ij}\}_{i\in I,j\in J}$ as its {\it matrix}. We say that a linear map $M:A\to B$ is {\it defined by its matrix} $\{M_{ij}\}_{i\in I,j\in J}$ if for any $a\in A$, $j\in J$ the series $c_j=\sum\limits_{i\in I}M_{ij}a_i$ converges absolutely and the sequence $c=\{c_j\}=Ma$.

{\bf Remark}\hspace{0.1 cm} It is important to note that not all maps of infinite-dimensional spaces are defined by their matrices. For example, for the space $l^\infty$ of bounded real sequences with the norm $$||\{a_n\}||=\sup\limits_n|a_n|,$$ by Hahn-Banach theorem (see \cite{Rudin} for a reference), we can define a bounded map from $l^\infty$ to ${\bf R}$, which sends all converging sequences, whose limit is zero, to zero and the constant sequence $\{1,\ldots,1,\ldots\}$ to $1$. Then the matrix of this map is zero, but not the map itself.
\bigskip

In the following we, as in the preceding Section, identify the index sets $I$ and $J$ with the set of natural numbers.

The following Lemma is a special case of a standard observation about maps of topological vector spaces.
\bl The map $M:A\to B$ is continuous iff  for any $b\in B^*$ there exists some $a'\in A^*$, $a'\geq0$, such that $\langle |Ma|,|b|\rangle \leq\langle a',|a|\rangle $ for all $a\in A$.
\el
{\bf Proof.}\hspace{0.01cm}
Assume $M$ is continuous and let $b\in B^*$.
Let $\epsilon>0$ be a number. By continuity of $M$ there exists $a'\in A^*$, $a'\geq 0$, and a number $\nu>0$ such that whenever $a\in A$ and $\langle a',|a|\rangle <\nu$ it holds that $\langle |Ma|,|b|\rangle <\epsilon$. Then $\langle a',|a|\rangle \leq\nu$ implies $\langle |Ma|,|b|\rangle \leq\epsilon$.

(Indeed, $\langle a',|a|\rangle \leq\nu$ implies $\langle a',|\lambda a|\rangle <\nu$ for $0<\lambda<1$, hence $\langle |M(\lambda a)|,|b|\rangle <\epsilon$. Taking the limit $\lambda\to 1$ we get the desired inequality.)

Rescaling $a'$ if necessary, we may choose even $\nu=\epsilon$.

Now, for  $a\in A$ let $\bar a=\nu\frac{a}{\langle a',|a|\rangle }$. Then $\langle a',|\bar a|\rangle \leq\nu$. Hence  $$\langle |Ma|,|b|\rangle =
\frac{1}{\nu}\langle a',|a|\rangle \langle |M\bar a|,|b|\rangle \leq\langle a',|a|\rangle \frac{\epsilon}{\nu}=\langle a',|a|\rangle .$$

Proof in the other direction is immediate. $\Box$
\bigskip

The following Lemma shows that continuous maps are defined by their matrices and the  transposes of these matrices define the adjoint maps.  Moreover, the absolute values of these matrix transposes induce well-defined maps as well.

\bl\label{map_matrix}
 Let the linear map $M:A\to B$ be continuous in normal topology. Then:

  (i) $M$ is defined by its matrix $\{M_{ij}\}_{i\in I,j\in J}$.

  (ii) There is also a map $M^*:B^*\to A^*$  defined by the transposed matrix $\{M^*_{ji}\}_{j\in J,i\in I}$, where $M^*_{ji}=M_{ij}$.

  (iii) There is a map $|M^*|:A\to B$ defined by the transposed matrix of absolute values $\{|M^*_{ji}|\}_{j\in J,i\in I}$.

   (iv) For any $a\in A$, $b\in B$, the series $\sum\limits_{i\in I,j\in J}a_iM_{ij}b_j$ converges absolutely to $\langle Ma,b\rangle $.

   (v) The map $M^*$ from (ii) is the adjoint of $M$, i.e. it satisfies $\langle Ma,b\rangle =\langle a,M^*b\rangle $.
   \el
{\bf Proof.}\hspace{0.01cm}
 (i) Consider $r_j\in B$, $j\in J$.

 By continuity of $M$ there exists $a'\in A^*$, $a'\geq 0$, such that $\langle |Ma|,r_j\rangle \leq|\langle |a|,a'\rangle |$ for any  $a\in A$. Pick $a\in A$, and let $\epsilon>0$ be a number. We know that the series in (\ref{pairing}), defining $\langle |a|,a'\rangle $ is absolutely converging.  So there exists $N$ such that $\sum\limits_{i=N}^\infty a'_i|a_i|<\epsilon$. Let $a^{\geq N}$ be the ``tail sequence'', coinciding with $a$ starting from the $i$th position, and zero elsewhere. We have $a^{\geq N}\in A$ and $\langle a',|a^{\geq N}|\rangle <\epsilon$. So $\langle |Ma^{\geq N}|, r_j\rangle <\epsilon$. Let $a^{\leq N}=a-a^{\geq N}\in S_I$. It follows that $\langle Ma^{\leq N},r_j\rangle \to\langle Ma,r_j\rangle $ as $N\to\infty$. But $\langle Ma^{\leq N},r_j\rangle =\sum\limits_{i=0}^N M_{ij}a_i$. So the series in the righthand side converges to $\langle Ma, r_j\rangle =(Ma)_j$.

To show that the convergence is absolute, consider the sequence $\tilde a\in A$, with $\tilde {a_i}=|a_i|\cdot|{M_{ij}}|/M_{ij}$.  Then $M_{ij}\tilde{a_i}= |M_{ij}||a_i|$, and by the identical reasoning $\sum\limits_{i=0}^\infty |M_{ij}||a_i|$ absolutely converges to $\langle M\tilde a, r_j\rangle <\infty$.

Thus $\sum\limits_{i=0}^\infty M_{ij}a_i$ converges absolutely to $(Ma)_j$, and $M$  is defined by its matrix.

(ii) Let $b\in B^*$. Again there exists an $a'\in A^*$, $a'\geq 0$, such that $\langle |Ma|,|b|\rangle \leq \langle |a|,a'\rangle $ for all $a\in A$.  Fix  an $i\in I$. We have $\langle |Me_i|,|b|\rangle \leq\langle e_i,a'\rangle $, that is
$\sum\limits_{j}|M_{ij}||b_j|\leq a_i'$.
  Thus the series $c_i=\sum\limits_{j}M_{ij}b_j$ converges absolutely, and $|c_i|\leq a_i'$. Then the sequence $c=\{c_i\}\in A^*$, since $a'\in A^*$, and $|c|\leq a'$. But $c$ is precisely the sequence $b$ multiplied by the transposed matrix $\{M^*_{ji}\}$. So the transposed matrix induces a well-defined map $B^*\to A^*$.

(iii) Same as (ii)

 (iv) For $b$ and $a'$ from the preceding paragraphs we have that the double series $\sum\limits_{i, j}|M_{ij}||b_j||a_i|\leq\sum\limits_{i}a_i'|a_i|=\langle a',|a|\rangle <\infty$.

Which means that the series $\sum\limits_{i,j}M_{ij}b_ja_i$ converges absolutely. Now $\sum\limits_{i}M_{ij}a_i$ for every $j\in J$ converges absolutely to $c_j$, where $c=Ma\in B$, and $\sum\limits_{j}c_jb_j$ converges absolutely to $\langle c,b\rangle $. So, for any $\epsilon>0$ and $n$ sufficiently large, $|\sum\limits_{j=0}^nc_jb_j-\langle c,b\rangle |< \frac{\epsilon}{2}$, and $|\sum\limits_{i=0}^nM_{ij}a_ib_j-\sum\limits_{j=0}^nc_jb_j|<\frac{\epsilon}{2}$, which means $|\sum\limits_{i=0}^nM_{ij}a_ib_j-\langle c,b\rangle |=|\sum\limits_{i=0}^nM_{ij}a_ib_j-\langle Ma,b\rangle |<\epsilon$.
This proves the claim.

(v) Follows immediately from (iv).
$\Box$
\bigskip

Let us say that a continuous map $M:A\to B$ is {\it continuously adjointable}, if the adjoint map $M^*$ is also continuous.
   The preceding Lemma implies the following
   \nb\label{continuous-adjointable-note} If $M:A\to B$ is continuously adjointable, then there is also the map $|M|: A\to B$ defined by the matrix $\{|M_{ij}|\}_{i\in I,j\in J}$ of absolute values.
\nbe
\bigskip

Let us show now that properties (i)-(v) from Lemma \ref{map_matrix} and the above Note in fact characterize continuously adjointable maps.
\bl\label{continuous-adjointable}
  In the setting as above, if  the matrix $\{M_{ij}\}_{i\in I, j\in J}$ defines a map from $A$ to $B$, and if its transposed matrix $\{M^*_{ji}\}_{j\in J, j\in I}$ defines a map from $B^*$ to $A^*$, and  similarly do the matrices $\{|M_{ij}|\}_{i\in I, j\in J}$, $\{|M^*_{ji}|\}_{j\in J, j\in I}$, and, finally, if for all $a\in A, b\in B^*$ the series $\sum\limits_{ij}a_iM_{ij}b_j$ converges absolutely,  then all defined maps are continuously adjointable.
\el
{\bf Proof.}\hspace{0.01cm}
Let us denote the maps in question as $M$, $M^*$, $|M|$ and $|M^*|$. Then $\langle Ma,b\rangle =\langle a,M^*b\rangle =\sum\limits_{ij}a_iM_{ij}b_j$, and
$\langle |M|a,b\rangle =\langle a,|M^*|b\rangle =\sum\limits_{ij}a_i|M_{ij}|b_j$, as follows from absolute convergence of the series.

Let  $b\in B$, $b\geq 0$. Put $a'=|M^*|b$, $a\geq 0$. Then for any $a\in A$ we have $\langle |Ma|,b\rangle \leq\langle |M||a|,b\rangle =\langle |a|,|M^*|b\rangle =\langle |a|,a'\rangle $. So for any $\epsilon>0$, whenever $\langle |a|,a'\rangle <\epsilon$ it holds that $\langle |Ma|,b\rangle <\epsilon$, so $M$ is  continuous.  Other maps are treated similarly.

Now since $M^*$ is continuous, and $M^*$ is indeed the adjoint of $M$, we conclude that $M$ is continuously adjointable. Same with other maps. $\Box$
\bigskip

The space of  continuously adjointable maps (as well as the space of all maps) of RSS is partially ordered: $M\geq N$ if $Ma\geq Na$ for all $a\ge 0$. Hence  the positive and negative parts $M^+, M^-$ of a continuously adjointable map, as well as  its absolute value $|M|$, are unambiguously defined, whenever they exist (with $M^+=\max(M,0)$ etc).

But it is easy to see that for maps defined by matrices the partial order is defined by the matrix order: $M\leq N$ iff $M_{ij}\leq N_{ij}$ for all $i\in I, j\in J$. And since continuously adjointable maps are defined by matrices, the partial order in their space is the matrix order. This yields the following.
\nb The absolute value  $|M|$ of a continuously adjointable map $M$ is the continuously adjointable map defined by the matrix $\{|M_{ij}|\}$.
\nbe
\bigskip

Similarly we find the positive and negative parts $M^+=\frac{1}{2}(|M|+M)\geq 0$, $M^-=\frac{1}{2}(|M|-M)\geq 0$ of a continuously adjointable $M$.

We define a  norm for continuously adjointable maps by
\be\label{map_norm}
||M||=\sup\limits_{||a||\leq 1, a\geq 0}||(|M|a)||.
\ee
Note that this is not the usual norm of a bounded map of Banach spaces.

We call a map $M:A \to B$ {\it regular} if it is continuously adjointable and $||M||<\infty$, i.e. $|M|$ is bounded.

\nb Regular maps compose.
\nbe
{\bf Proof.}\hspace{0.01cm} Let $A,B,C$ be RSS and the maps $M:A\to B$, $N: B\to C$ be regular. We
have  decompositions $M=M^+-M^-$, $N=N^+-N^-$ of $M$ and $N$ into positive and negative parts. Then $NM=K^+-K^-$, where $K^+=N^+M^++N^-M^-$, $K^-=(N^+M^-+N^-M^+)$, $K^+,K^-\geq 0$. It follows that $K^+\geq NM$, hence $K^+\geq max(NM,0)=(NM)^+$, and, similarly $K^-\geq (NM)^-$. So $|N||M|=K^++K^-\geq|NM|$. Since $|N||M|$, being the composition of bounded maps, is bounded, it follows that $|NM|$ is bounded.
\bigskip

Since the identity maps of RSS are obviously regular, the following is well-defined.
\bd The category {\bf RSS} of rigged sequences spaces has rigged sequences spaces as objects and regular maps of norm less than or equal to 1 as morphisms.
\ed

We call regular maps of norm less or equal to $1$ {\it contracting}.

 Thus we have given an intrinsic, coordinate-free definition of morphisms.

 On the other hand, as Lemmas above show, we can equivalently  describe morphisms as  bounded adjointable maps defined by matrices.

Seen as matrices, regular maps have the following characterization.
\bl\label{matrix-map}
Given RSS $S_I\subseteq A\subseteq S_I^*$, $S_J\subseteq B\subseteq S_J^*$, if the matrix $\{M_{ij}\}_{i\in I,j\in J}$, satisfies $$\sup\{\sum\limits_{i\in I,J\in J}a_i|M_{ij}|b_j~~|a\in A, b\in B^*, a,b\geq 0, ||a||,||b||\leq 1\}=K<\infty,$$ then $\{M_{ij}\}$ defines a continuous adjointable map $M:A\to B$ with $||M||=K$.
\el
{\bf Proof.}\hspace{0.01cm} It is easy to see that the matrix satisfies conditions of Lemma \ref{continuous-adjointable}, and the claim follows. $\Box$
\bigskip

 Finitely let us note that any map between spaces of finite sequences is defined by its matrix. The following routine  Note, says when such a map extends to a map of RSS.
\nb\label{map_norm_finit} Given RSS $S_i\subseteq A_i\subseteq S_i^*$, $i=1,2$,  a  linear map  $M:S_1\to S_2$  extends continuously to a regular map from $A_1$ to $A_2$ iff the supremum
$$\sup\limits_{||a||=1, a\in S_1}||(|M|a)||=\sup\limits_{||a_i||=1,  a_i\in S_i, i=1,2}|\langle a_1,|M|a_2\rangle |$$ is finite, in which case this supremum coincides with  norm  (\ref{map_norm}) of $M$.
\nbe
\bigskip

{\bf Remark}\hspace{0.1 cm}
The reader familiar with probabilistic coherence spaces (see \cite{DanosErhahrdPCS}) will observe that, under identification of RSS with Banach spaces associated to PCS, morphisms of PCS are precisely positive morphisms of RSS, i.e. regular maps which are positive. It is worth noting also that an adjointable positive map is automatically continuous  and continuously adjointable, hence morphisms of PCS can be characterized simply as bounded positive adjointable maps.

\subsubsection{A non-morphism}
It is reasonable to ask if our definition of morphisms is not redundant, i.e. if  adjointable bounded  non-regular maps exist. The answer is yes, and the following example is based on the one given for PCS in \cite{DanosErhahrdPCS}, slightly rephrased.

Let $W$ be the set of finite binary strings of even length. For a binary string $w$, let $|w|$ be the length of $w$, and let $w_i$ be the $i$-th bit of $w$. Also, let
$$D(w)=|\{\mbox{number of ones in }w\}-\{\mbox{number of zeros in }w\}|.$$
Consider the RSS $A$ with the index set ${\bf N}$ and the $l^\infty$-norm given by
$$||a||=\sup\limits_i|a_i|$$
and  $B$ with the index set $W$ and the $l^1$-norm given by
$$||b||=\sum\limits_w|b_w|.$$
(It is clear from Note \ref{sup_over_finite_norms} that these norms indeed define RSS).

Let the linear operator $L:A\to B$ be given by
$$Le_i=\sum\limits_{|w|\geq i}\frac{(-1)^{w_i}}{|w|^22^{|w|}}e_w,$$
extended to all sequences in $A$ by linearity: $La=\sum\limits_ia_iLe_i$. (As usual, $e_i$ denotes the sequence with $1$ at the $i$-th position and zeros elsewhere, similarly with $e_w$.)
It is easy to see that the sequence $La$ is well defined, we need to check that it belongs to $B$, i.e. that its $l^1$-norm is finite.

We have for any sequence $a\in A$ that
$$(La)_w=\sum\limits_{i\leq|w|}\frac{(-1)^{w_i}a_i}{|w|^22^{|w|}},$$ and it is easy to see that  $$|(La)_w|\leq \frac{D(w)}{|w|^22^{|w|}}||a||.$$ Then
$$\sum\limits_{|w|=2n}|(La)_w|\leq\frac{1}{4n^24^{n}}||a||\sum\limits_{|w|=2n}D(w)=$$
$$=\frac{1}{4n^24^{n}}||a||\sum\limits_{k=0}^{2n}\frac{(2n)!}{(2n-k)!k!}2(n-k).$$
After some combinatorics and an application of Stirling formula we get the estimate
$$\sum\limits_{k=0}^{2n}\frac{(2n)!}{(2n-k)!k!}2(n-k)=\frac{(2n)!}{(n!)^2}\leq Cn^{\frac{1}{2}}4^n,$$
where $C$ is some constant independent of $n$ (see
\cite{DanosErhahrdPCS} for a derivation).
Hence $$\sum\limits_{|w|=2n}|(La)_w|\leq C\frac{n^{\frac{1}{2}}4^n}{4n^24^{n}}||a||=C\frac{||a||}{4n^{\frac{3}{2}}},$$
and
$$||Lw||\leq K||a||,$$ where
 $K=C\sum\limits_n\frac{1}{4n^{\frac{3}{2}}}<\infty$.
 Thus $Lw\in B$, moreover, $L$ is bounded. The reader is invited to check that $L$ has a well-defined adjoint $L^*:B^*\to A^*$.

On the other hand, the absolute value $|L|$ of $L$, if it exists, must map $e_i$ to $\sum\limits_{|w|\geq i}\frac{1}{|w|^22^{|w|}}e_w$ and extend to other sequences by linearity. In particular, for the sequence $\Omega=\{1,1,\ldots,1,\ldots\}$ we must have $$(|L|\Omega)_w=\sum\limits_{i\leq|w|}\frac{1}{|w|^22^{|w|}}=\frac{1}{|w|2^{|w|}},$$ and it is immediate to compute that $||(|L|\Omega)||=\infty$. Hence $|L|$ not only is unbounded, but even does not take values in $B$. It follows that $L$ is non-regular.

\subsection{Products and {*}-autonomous structure}
The category {\bf RSS} has products and coproducts as  any reasonable category of normed spaces.
For  RSS $A,B$ their product $A\times B$, respectively, coproduct $A\oplus B$ is the vector space $A\times B$ equipped with the norm $||(u,v)||=max(||u||,||v||)$, respectively, $||(u,v)||=||u||+||v||$. The index set for $A\times B$ and $A\oplus B$ is the disjoint union of index sets for $A$ and $B$. Spelling out the details is absolutely routine and is left to the reader.
\nb In {\bf RSS}, for any objects $A,B$ their product $A\times B$ and coproduct $A\oplus B$ exist, and are dual: $A\times B=(A^*\oplus B^*)^*$. The neutral object for both operations is the zero-space $\{0\}$.
\nbe

We now describe briefly the monoidal closed structure, which also seems quite clear.

For RSS $S_i\subseteq A_i\subseteq S_i^*$, $i=1,2$, the space of  regular maps is, by Lemma \ref{matrix-map}, identified as a subspace of matrices, i.e. of sequences on the set $I_1\times I_2$. The norm $||.||^*$ on the subspace  of finite sequences is defined by  formula (\ref{map_norm}) or, equivalently by Lemma \ref{matrix-map}, which defines in its turn the dual norm $||.||^*$ by (\ref{dual_norm}). It is routine to check then that the space of regular maps consists of matrices of finite norm $||.||$ extended to all sequences by (\ref{norm_ext}). So this is an RSS, which we denote as $A_1\multimap A_2$, on the index set $I_1\times I_2$ (this is especially evident from Note \ref{map_norm_finit}).

Identifying the base field ${\bf R}$ with the self-dual RSS on  the one-element index set and denoting it as ${\bf 1}=\bot$, we check that for any $A\in {\bf RSS}$ the dual $A^*=A\multimap\bot$.

Having defined the maps space, it is routine to describe the cotensor and tensor products by duality as $A\wp B=A^*\multimap B$, $A\otimes B=(A^*\wp B^*)^*$. We leave it to the reader. We just state the summarizing theorem.
\bt
The category ${\bf RSS}$ is $*$-autonomous, with the internal hom-space $A\multimap B$ being the space of regular maps between $A$ and $B$ with the norm given in (\ref{map_norm}) and with the dualizing object $\bot={\bf R}$.
\et

\section{Exponentials}
We proceed now to discussing exponentials.  In fact, the main interest of the current work is that the category $\bf RSS$ has {\it free} exponentials.

Recall that the linear logic exponential modality $?$ (respectively $!$) corresponds, on the semantic side, to an endofunctor sending  every object to a {\it commutative monoid} with respect to the monoidal structure $\wp$ (respectively to a {\it commutative comonoid} with respect to the monoidal structure $\otimes$). Let us spell this out.

 Let $\bf K$ be a symmetric monoidal category. We denote the monoidal product as $\wp$ and the monoidal unit, as $\bot$. A monoid in $({\bf K},\wp,\bot)$  is an object $X$ together with two {\it monoidal maps}  $$m:X\wp X\to X,\mbox{ }u:\bot\to X,$$  respectively {\it multiplication} and  {\it unit}. The monoidal maps must satisfy certain axioms corresponding to associativity of multiplication and neutrality of unit. In the case of the exponential object $?A$ multiplication and unit correspond to possibility of duplicating or discarding a hypothesis, that is, to rules of contraction and weakening.

 The monoid $X$ is {\it commutative}, if the multiplication $m$ commutes with the symmetry map. A morphism of monoids in $\bf K$ is a ${\bf K}$-map preserving multiplication and unit; this condition is expressed by a couple of commutative diagrams, which we do not give here. Comonoids, commutative comonoids and comonoid morphisms are defined dually, see \cite{Mellies_categorical_semantics} for details and discussion.

 A monoid $?A$ is {\it free} over or {\it freely generated } by the object $A$ if there exists a map $$\eta_A:A\to?A,$$ such that for any morphism $$f:A\to X,$$ where $X$ is a monoid, there exists a unique monoid morphism $$\hat f:?A\to X$$ such that $$\hat f\circ\eta_A=f.$$ By duality one obtains a definition of a {\it freely generated comonoid} $!A$. It was found in \cite{Lafont} that, if in a $*$-autonomous category ${\bf K}$ there is a  free comonoid $!A$ (dually, free monoid $?A$) for every object $A$, then this category is a model of full propositional linear logic with exponentials interpreted as free constructions. Such a category is called {\it Lafont category}, again, see \cite{Mellies_categorical_semantics} for more details.

It turns out that $\bf RSS$ is a Lafont category, that is, for any RSS $A$ there exist the   monoid $?A$ and, by duality, the  comonoid $!A$, freely generated by $A$.

\subsection{Exponentials in ${\bf RSS}$}
We now give an explicit construction of the RSS $?A$ and $!A$ and postpone the proof that these are free and, consequently, give a model of linear logic exponentials.

Let $A$ be an RSS with the index set $I$. For any $n\geq 0$ consider the RSS $\wp^nA$ of $n$-linear functionals on $A^*$. It is the RSS with the index set $I^n$ and  the norm given by
\be\label{n-linear norm}
||a||=\sup\limits_{||x_i||\leq 1,x_i\in A^*,i=1,\ldots,n}|a|(|x_1|,\ldots, |x_n|).
\ee
Now, seen as a Banach space, $\wp^nA$ contains the subspace $S^nA$ of {\it symmetric $n$-linear functionals} $a$, satisfying $$a(x_1,\ldots, x_n)=a(x_{\pi(1)},\ldots,x_{\pi(n)})$$ for any permutation $\pi\in S_n$. The space $S^nA$ itself has a description as an RSS.

Indeed, let  $M_n(I)$ be the set of all cardinality $n$  multisets in $I$. For convenience, let us identify, as usual, $I$ with a subset of natural numbers. Then any element $a\in S_{M_n(I)}$ induces a symmetric functional on $A^*$, defined by
 $$a(x_1,\ldots,x_n)=\frac{1}{n!}\sum\limits_{m\in M_n(I)}(a_m\cdot\sum\limits_{\substack{i_1<\ldots<i_n,\\ i_1,\ldots,i_n\in m}}(x_1)_{i_1}\ldots(x_n)_{i_n}),$$
  where $x_1,\ldots,x_n\in A^*$.
  (For clarity: $(x_k)_{i_j}$ is the $i_j$-th element of the sequence $x_k$. Note that each monomial in the above sum occurs $n!$ times because of different orderings $i_1<\ldots<i_n$.)

  Norm (\ref{n-linear norm}) on $S_{M_n(I)}$ gives rise to an RSS of symmetric $n$-linear functionals, which can be identified with $S^nA$.

  Any element $a\in S^nA$ also gives rise to a {\it homogenous $n$-th degree polynomial} on $A^*$, which we, abusing the notation, denote by the same letter:
  $$a(x)=\frac{a(x,\ldots,x)}{n!}.$$
  Now we define a {\it  power series} on $A^*$ as a sequence $a=\{a_0,\ldots,a_n,\ldots\}$, where $a_n\in S^nA$. The homogenous polynomials $a_0,a_1,\ldots$  are the {\it coefficients} of $a$.

  The power series $a$  {\it converges absolutely} at the point $x\in A^*$,  if the series
  $\sum\limits_n|a_n|(x)$ converges absolutely. The power series is a {\it polynomial} if for $N$ sufficiently large all $a_n$ with $n>N$ are zero.

 The {\it absolute value} $|a|$ of the power series $a$ is the power series with the coefficients $|a_0|,\ldots,|a_n|,\ldots$. Any power series $a$ induces two partial functions on $A^*$, which we also denote as $a$ and $|a|$, with the usual abuse of notation. These are   defined at the point $x\in A^*$, iff $a$ converges absolutely at $x$, and are given by
 $$a(x)=\sum\limits_na_n(x),\mbox{ }|a|(x)=\sum\limits_n|a_n|(x).$$

  The  space $?A$ is the space of power series on $A^*$ that absolutely converge at every point $x\in A^*$ with $||x||\leq 1$ and are bounded in the norm
  \be\label{power_series_norm}
  ||a||=\sup\limits_{x\in A^*,||x||\leq 1}|a|(|x|).
  \ee

  It is clear from the above discussion that any power series $a$ can be encoded as a sequence on the index set $M(I)$ of finite multisets on $I$. This sequence is finite, if $a$ is a finitely supported polynomial. Otherwise, it is easy to see that $||a||$ is the supremum of $||p||$ over the set of all finitely supported positive polynomials $p\leq|a|$.
  It follows from Note \ref{sup_over_finite_norms} that $?A$ is the RSS defined on the index set $M(I)$ by norm (\ref{power_series_norm}).

  \bd For any RSS $S_I\subseteq A\subseteq S_I^*$ the space $?A$ is the RSS defined on the index set $M(I)$ of finite multisets on $I$ by norm ({\ref{power_series_norm}}).

  The RSS $!A$ is the dual of $?A^*$.
  \ed

  The space $?A$ has a natural interpretation as a space of functions on the unit ball of $A^*$. These functions are given by absolutely convergent power series and it is reasonable to call them real analytic. One should keep in mind however that real analytic functions of such a form, in general, are not continuous in any sense at the boundary of the unit ball, as can be seen already in the finite-dimensional case.

  The RSS $!A$ in its turn has an interpretation as a space of real analytic distributions supported in the unit ball. Note in particular that $!A$ contains all {\it delta-functions} $\delta_x$ for $x\in A$, $||A||\leq 1$, defined on $?A^*$ by
  \be\label{delta}
  \langle a,\delta_x\rangle=a(x).
   \ee
   Indeed, any such functional has a formal expansion $\delta_x=\sum\limits_nx^{\otimes n}$, where $n$-th summand belongs to $S^nA^*$. Hence, $n$-th summand   can be encoded as a sequence on $M_n(I)$, where $I$ is the index set of $A$, and the whole $\delta_x$ can be encoded as a sequence on $M(I)$. It remains to observe that the $!A$-norm of this sequence is finite.

  It turns out that the RSS $?A$ is a free monoid in ${\bf RSS}$ generated by $A$, and, dually, $!A$ is a freely generated comonoid, hence, by \cite{Lafont} they give a model of linear logic exponential connectives.
  We postpone  a proof and first briefly describe the model  without going into details.

  The unit $u_A:{\bf 1}\to ?A$ for the monoid $?A$ is the map, sending a number to the corresponding constant function on $A^*$. As for the multiplication  $m_A:?A\wp?A\to ?A$, it is given by  the {\it diagonalization} map. The space $?A\wp?A$ consists of double power series, corresponding to  functions of two variables on $A^*$, and the diagonalization map sends the function $f$ of two variables on $A^*$ to the function $m_af$ defined on the unit ball of $A^*$ by $$(m_af)(x)=f(x,x).$$ The comonoidal structure on $!A$ is dual.

  The action of the functor $?$ on morphisms is as follows. For the morphism $$L:A\to B,$$ the morphism $$?L:?A\to?B$$ sends a function $f\in ?A$ to the function $$?Lf=f\circ `L^*,$$ which belongs to $?B$.

  Finally, we describe  the important  monadic maps
  $$\eta_A:A\to ?A, \mbox{ }\mu_A:??A\to ?A,$$ used to interpret principles of dereliction and digging. The map $\eta_A$ sends an element of $A$ to the corresponding linear function on $A^*$, and $\mu_A$ is defined by
  \be\label{digging}
  (\mu_Af)(x)= f(\delta_x).
  \ee

  The above operations are indeed well-defined and satisfy all necessary properties for a model of linear logic,
  which follows from the theorem we are going to prove next: namely that the above construction of monoid $?A$ and comonoid $!A$ is free.

  {\bf Remark}\hspace{0.1 cm} At this point it seems reasonable to discuss  differences with the seminal work \cite{Girard99coherentbanach} on coherent Banach spaces, where exponentials do not work. Recall that in \cite{Girard99coherentbanach} the space $?A$ was interpreted as the Banach space of analytic functions on the {\it open} unit ball of $A^*$, bounded in the sup-norm $||f||=\sup\limits_{||x||< 1}|f(x)|$. The problem with such an interpretation is that analytic functions, in general, have no reasonable extension from the open ball to the  closure. On the other hand, $\delta$-functions (\ref{delta}) live precisely at the boundary of  the unit ball in $!A$, and this eventually leads to troubles with various constructions, in particular with interpretation (\ref{digging}) of digging.

  A banal observation is that, in our model, norm (\ref{power_series_norm}) is  not the sup-norm, except for the case of power series with positive coefficients. A more sensible comment is that our functions are dominated by positive power series  converging to uniformly bounded sums everywhere on the init ball. Therefore they have  canonical extensions to the boundary, although these may be discontinuous. Apparently, a crucial role is played by positivity.

\subsection{Abstract construction of free monoids}
In \cite{Mellies_free_exp} it was found that, under certain conditions on a $*$-autonomous category, its objects freely generate  comonoids (respectively, monoids), hence such a category  carries an interpretation of exponentials as free constructions. It turns out that the category $\bf RSS$ falls in this class.

We first describe the abstract construction of a free monoid given in \cite{Mellies_free_exp}. More pedantically, we describe a dual version, because  in \cite{Mellies_free_exp} Melli\'es et al. constructed a freely generated comonoid, rather than a monoid.

Consider the {\it free copointed object} $A^\bullet$ on $A$, $A^\bullet=A\oplus\bot$.

For each $n$ consider the object $$\wp^n A^\bullet=A^\bullet\wp\ldots\wp A^\bullet\mbox{ }n\mbox{ times}.$$
The permutation group $S_n$ acts on $\wp^n A^\bullet$ by endomorphisms in an obvious way: any permutation $\pi\in S_n$ induces the natural symmetry transformation $$\tau_\pi:A_1\wp\ldots\wp A_n\to A_{\pi(1)}\wp\ldots\wp A_{\pi(n)}$$  by iteration of the canonical symmetry transformation $\tau:X\wp Y\to Y\wp X$.

Recall that a {\it coequalizer} of a group $G$ of endomorphisms   acting on the object $X$ is an object $E$ together with the map $$\rm{coeq}:X\to E,$$ satisfying the following universal property. For any $g\in G$ it holds that $$\rm{coeq}\circ g=\rm{coeq},$$ and for any other object $Y$ and map $m:X\to Y$ satisfying $m\circ g=m$ there exists a unique map $$u:E\to Y,$$ such that
$$m=u\circ\rm{coeq}.$$
(Equalizer is defined dually.)

 Now the object $A_{\leq n}$ is defined as the coequalizer of the group $S_n$ action on $\wp^n A^\bullet$, {\it assuming it exists}.  Thus  $A_{\leq n}$ is the $n$-th {\it symmetric cotensor power} of $A^\bullet$.

The canonical injection $\bot\to A^\bullet$ induces morphisms
$$\wp^nA^\bullet\cong(\wp^nA^\bullet)\wp\bot\to\wp^{n+1} A^\bullet,$$
 which give rise to the commutative diagrams
\begin{diagram}
   {\wp^{n}A^\bullet}      &\rTo &{\wp^{n+1}A^\bullet}&\rTo  & {A_{\leq n+1}}      \\
   \dTo_{\tau_\pi}&          &   \dTo_{\tau_\pi\wp\Id}         &\ruTo         &\\
         {\wp^nA^\bullet}       &\rTo    & {\wp^{n+1}A^\bullet}   &              &
\end{diagram}
for all $\pi\in S_n$. Hence
 by the universal property of coequalizers we have morphisms $A_{\leq n}\to A_{\leq n+1}$ and, in a similar fashion, morphisms
\be\label{i_mn_def}
 i_{mn}:A_{\leq m}\to A_{\leq n}\mbox{ for }m\leq n.
 \ee

Thus, assuming that all relevant coequalizers exist,  we get the sequence of maps
\be\label{seq. system}
\bot\cong A_{\leq 0}\to A_{\leq 1}\to\cdots\to A_{\leq n}\to\cdots.
\ee

Recall that a {\it sequential colimit} of the sequence of maps
\be\label{seq_colim_def}f_i:X_i\to X_{i+1},i=0,1,\ldots
\ee
is an object $X$ together with the sequence of maps $$g_i:X_i\to X$$ satisfying the following universal property.
For all $i$ it holds that $$g_{i+1}\circ f_i=g_i,$$ and for any other object $Y$ and maps $h_i:X_i\to Y$ satisfying $h_{i+1}\circ f_i=h_i$, $i=0,1,\ldots$, there exists a unique map $$h:X\to Y$$ such that $$h_i=h\circ g_i,\mbox{ }i=0,1,\ldots$$ (Sequential limit is defined dually.)

Let us say that the sequential colimit $X$ of system (\ref{seq_colim_def}) {\it commutes with cotensor product}, if for any object $Y$ the object $Y\wp X$ is a sequential colimit of the system $\Id\wp f_i:Y\wp X_i\to Y\wp X_{i+1}$.

The desired monoid  is the sequential colimit $A_\infty$ of  system (\ref{seq. system}), {\it assuming it exist and commuted with cotensor product}.

\bt\label{Mellies}(Melli\'es et al.\cite{Mellies_free_exp}) In the setting as above, if all relevant coequalizers exist, and a sequential colimit exists and commutes with cotensor product,  then the constructed object $A_\infty$ is a free commutative monoid generated by $A$.
\et

\subsection{Multiaffine functionals}
We are now going to apply the general construction from \cite{Mellies_free_exp} described above to the special setting of $
\bf RSS$.  In this Section we construct the objects $A_{\leq n}$.

Recall that a functional $a$ on the vector space $V$ is {\it affine}, if there exists a linear functional $a_0$ such that $a(x)-a(y)=a_0(x-y)$ for all $x,y\in V$. Equivalently, $a$ is affine, if it can be written as
\be\label{affine}
a(x)=a_0(x)+a_1,
\ee
  where $a_0\in A$ is linear, and $a_1\in \bf R$ is an additive constant.

Then, for a given rigged sequences space $A$, the copointed object $A^\bullet\cong A\oplus\bot=A\oplus{\bf R}$ has a natural interpretation as the space of affine functionals on $A^*$ of  form (\ref{affine}), equipped with the norm
\be
||a||=\sup\limits_{||x||\leq 1}|a|(|x|),
\ee
the absolute value of $a$ being defined by
\be
|a|(x)=|a_0|(x)+|a_1|.
\ee

Let us say that a functional of several variables is {\it multiaffine}, if it is affine in every variable, when other variables are fixed. We are going to show that the coequalizers $A_{\leq n}$ have a similar interpretation as spaces of symmetric {\it multiaffine} functionals of $n$ variables on $A^*$.

The space $A_{\leq n}$ is the space $S^nA^\bullet$ of symmetric $n$-linear functionals on $(A^\bullet)^*=A^*\times{\bf 1}=A^*\times{\bf R}$, bounded in the corresponding norm, which is defined as in (\ref{n-linear norm}).
The coequalizer map $Sym_n:\wp^nA^\bullet\to A_{\leq n}$
is averaging over permutations:
\be\label{symmetrization}
Sym_n(a)(x_1,\ldots, x_n)=\frac{1}{n!}\sum\limits_{\pi\in S_n}a(x_{\pi(1)},\ldots,x_{\pi(n)}).
\ee

Recall that the index set for the RSS $A^\bullet$ is $I^\bullet =I\cup\{*\}$, where $I$ is the index set for $A$, and the index set for $A_{\leq n}=S^nA^\bullet$ is the set $M_n(I^\bullet)$ of cardinality $n$ multisets on $I^\bullet$. Then $M_n(I^\bullet)$ can be identified with the set $M_{\leq n}(I)$ of cardinality less than or equal to $n$ multisets on $I$. Namely, we identify a cardinality $k\leq n$ multiset $m$ on $I$ with the multiset on $I^\bullet$ containing the same elements with the same multiplicities as $m$ and, in addition, the element $*$ with multiplicity $n-k$. This suggests that elements of RSS $A_{\leq n}$ can be identified with sequences of symmetric $k$-linear functionals on $A^*$, where $k\leq n$, and this is indeed the case.

Any symmetric $n$-linear functional $a$ on $A^*\times{\bf R}$  gives rise to the $n$-ary functional
\be\label{linear_to_affine}
x_1,\ldots,x_n\mapsto a((x_1,1),\ldots,(x_n,1))
 \ee
 on $A^*$, which is immediately seen to be affine in each variable.

On the other hand any symmetric multiaffine functional $a$ of  $n$ variables on $A^*$ induces the symmetric functional
\be\label{affine_to_linear}
(x_1,t_1),\ldots,(x_n,t_n)\mapsto\lim\limits_{s_1\to t_1}\ldots\lim\limits_{s_n\to t_n}a(\frac{x_1}{s_1},\ldots,\frac{x_n}{s_n})s_1\ldots s_n
\ee
on $A^*\times{\bf R}$. The latter functional is linear in each variable, which is evident from decomposition of an affine functional into the linear and the constant parts.

Formulas (\ref{affine_to_linear}), (\ref{linear_to_affine}) establish a one-to-one correspondence.

Also, any symmetric $n$-linear functional $a$ on $A\times{\bf R}$ gives rise to the $(n+1)$-tuple $\{a_0,\ldots,a_n\}$, where each $a_k$ is a symmetric $k$-linear functionals on $A^*$, defined by
\be\label{functional_to_tuple}
a_k(x_{1},\ldots,x_{k})=a((x_{1},0),\ldots,(x_{k},0),(0,1),\ldots,(0,1)).
\ee

Conversely any $(n+1)$-tuple $\{a_0,\ldots,a_n\}$, where each $a_k$ is a symmetric $k$-linear functional on $A^*$, gives rise to the symmetric $n$-linear functional $a$ on $A^*\times{\bf R}$ defined by
\be\label{tuple_to_functional}
a((x_1,t_1),\ldots,(x_n,t_n))=\sum\limits_{k=0}^n\sum\limits_{\pi\in S_n}\frac{a_k(x_{\pi(1)},\ldots,x_{\pi(k)})t_{\pi(k+1)}\ldots t_{\pi(n)}}{k!(n-k)!}.
\ee

Again, it is easy to check that formulas (\ref{functional_to_tuple}), (\ref{tuple_to_functional}) establish a one-to-one correspondence $a\mapsto\{a_0,\ldots,a_n\}$.

\nb\label{tuple_absolute_values}
In the above correspondence,  if the symmetric $n$-linear functional $a$ on $A^*\times{\bf R}$ corresponds to the $n$-tuple $\{a_0,\ldots,a_n\}$, then $|a|$ corresponds to $\{|a_0|,\ldots,|a_n|\}$.

Also, $a\geq0$ iff $a_i\geq0$ for all $i=0,\ldots,n$.
\nbe
{\bf Proof.}\hspace{0.01cm} The first claim is immediate from (\ref{functional_to_tuple}). The second claim follows from  the first one and the fact that $a\geq 0$ iff $a=|a|$.
\bigskip

Combining the two above correspondences,  we get that any $n$-ary multiaffine functional $a$ on $A^*$ can be encoded as the $(n+1)$-tuple $\{a_0,\ldots,a_n\}$ of symmetric multilinear functionals on $A^*$. And combining formulas (\ref{tuple_to_functional}) and (\ref{linear_to_affine}), we obtain an expression for the multiaffine $a$ in terms of the corresponding tuple:
\be\label{multiaffine}
a(x_1,\ldots,x_n)=\sum\limits_{k=0}^n\sum\limits_{1\leq i_1<\ldots<i_k\leq n}a_k(x_{i_1},\ldots,x_{i_k}).
\ee
Let us define   partial  order on symmetric $n$-ary multiaffine functionals by
\be\label{affine_order}
a\geq b{ \mbox{ iff } }a_i\geq b_i, i=0,\ldots,n.
\ee
Then the absolute value $|a|$ of symmetric multiaffine $a$ is defined by the tuple $\{|a_0|,\ldots,|a_n|\}$. It follows from Note \ref{tuple_absolute_values} that, with such a definition, the one-to-one correspondence between symmetric multiaffine functionals on $A^*$ and symmetric multilinear ones on $A^*\times{\bf R}$ is order-preserving.

Finally, observe that, for a non-negative multilinear functional in variables $(x_1,t_1),\ldots,(x_n,t_n)$ with $x_i\in A^*$, $t_i\in {\bf R}$, supremum on the unit ball of $A^*\times {\bf R}$ is determined by vectors with $t_1=\ldots=t_n=1$. It follows then from formula (\ref{linear_to_affine}) that the norm of a symmetric $n$-ary multiaffine functional $a$ on $A^*$, defined by
 \be\label{multiaffine_norm}
||a||=\sup\limits_{||x_i||\leq 1, i=1,\ldots, n}|a|(|x_1|,\ldots,|x_n|),
\ee
coincides with norm of the corresponding symmetric $n$-linear functional on $A^*\times{\bf R}$.

 Thus we obtain that the Banach space $A_{\leq n}$ is isomorphic to  the space of multiaffine functionals  in $n$ variables on $A^*$ that are bounded in norm $(\ref{multiaffine_norm})$. If we agree to encode symmetric multiaffine functionals as tuples of symmetric multilinear functionals, it is easy to see that the latter space becomes an RSS with the index set $M_{\leq n}(I)$.  Observe also  that the isomorphism between  $A_{\leq n}$ and the corresponding Banach space of multiaffine functionals  is consistent with the previously discussed identification of the index sets $M_{\leq n}(I)$
 and $M_n(I^\bullet)$.  From now on we  interpret $A_{\leq n}$  as an RSS of multiaffine functionals with the index set $M_{\leq n}(I)$.

 In this interpretation the coequalizer  map has the same form (\ref{symmetrization}). Let us compute maps $i_{mn}:A_{\leq m}\to A_{\leq n}$  for $m\leq n$, discussed in the preceding subsection (see (\ref{i_mn_def})).

 The canonical maps $\wp^m A^\bullet\to\wp^n A^\bullet$ for $m\leq n$  introduce fake dependencies: the $m$-ary functional $a$ on $A^\bullet$ is sent to the $n$-ary functional
$$(x_1,t_1),\ldots, (x_n,t_n)\mapsto a((x_1,t_1),\ldots, (x_m,t_m))t_{m+1}\ldots t_n.$$
Combining this with the symmetrization and using our identification with multiaffine functionals on $A^*$, we obtain  the  maps $i_{mn}$, defined on symmetric multiaffine functionals    by
\be\label{poly_to_poly}
i_{mn}(a)(x_1,\ldots,x_{n})=\frac{1}{n!}\sum\limits_{\pi\in S_{n}}a(x_{\pi(1)},\ldots,x_{\pi(m)}).
\ee

 Using formula (\ref{multiaffine}), we compute the image $i_{mn}a$ of $a\in A_{\leq m}$, $m\leq n$, explicitly as
\be\label{i_mn}
(i_{mn}a)(x_1,\ldots,x_n)=\frac{1}{n!}\sum\limits_{k=0}^m(\frac{m!(n-k)!}{(m-k)!}\sum\limits_{1\leq i_1<\ldots<i_k\leq n}a_k(x_{i_1},\ldots,x_{i_k})).
\ee

Indeed, for a $k$-tuple $i_1<\ldots<i_k$ of integers between $1$ and $n$, and a $k$-tuple $j_1<\ldots<j_k$ of integers between $1$ and $m$ there are $k!(n-k)!$ permutations that map the latter to the former, and there are $\frac{m!}{k!(m-k)!}$ $k$-tuples between  $1$ and $m$. Hence the expression $a_k(x_{i_1},\ldots,x_{i_k})$ occurs in the expansion of $i_{mn}a$ exactly $\frac{m!(n-k)!}{(m-k)!}$ times.

The maps $\{i_{mn}\}_{m\leq n}$ constitute the sequence, whose colimit corresponds to the $?$-modality. We need to show that this colimit exists.

\subsection{Sequential limits and colimits}
We will consider  the following situation, where  colimits
 are guaranteed to exist.

Let $I_1\subseteq\ldots\subseteq I_n\subseteq\ldots$ be a sequence of index sets, and $A_1\ldots,A_n\ldots$ be a sequence of RSS, $S_i\subseteq A_i\subseteq S_i^*$, where $S_i$ is the space of finite
sequences on $I_i$, and $S_i^*$ is the space of all sequences on $I_i$. Inclusions of the index sets induce linear maps $f_{ij}:S^*_i\to S^*_j$ between corresponding sequence spaces  for $j\geq i$, if we identify elements of $I_i$ with basis vectors of $S_i^*$. We assume that these maps restrict to regular maps $f_{ij}:A_i\to A_j$ of norm less or equal to $1$, i.e. to morphisms in ${\bf RSS}$. Since $f_{jk}\circ f_{ki}=f_{ji}$ for $i\leq k\leq j$ this gives us a commutative diagram in ${\bf RSS}$.

Let us denote the norm on $A_i$ as $||.||_i$. Let $I=\bigcup\limits_i I_i$ and let $S$ be the space of finitely supported sequences on $I$. For any $a\in S$ it holds that $a\in S_i\subseteq A_i$ for $i$ sufficiently large, and since all maps $f_{ij}$ are contracting it follows that the sequence $\{||a||_i\}$ is non-decreasing and bounded from below by zero. Hence it converges. Define
\be\label{norm_colimit}
||a||=\lim\limits_{i\to\infty}||a||_i.
\ee
We assume that for all non-zero $a\in A$ the limit is non-zero. In fact, it is sufficient to require this for all basis vectors $\{e_i\}$, ${i\in I}$, where $e_ i\in S$ is the corresponding sequence, with zeros everywhere except the $i$-th position. Then (\ref{norm_colimit}) defines a norm on $S$ indeed.

This data is sufficient to define a rigged sequence space. As usual, let $S^*$ be the space of all sequences on $I$, and set $A$ to be the subspace of sequences for which  norm (\ref{norm_colimit}) is finite. We will show shortly that $A$ is the desired colimit.

Let us describe also the dual space, i.e. the {\it limit} of the dual diagram. On the dual side we have the system of projections $f_{ij}^*:A_j^*\to A_i^*$ of dual RSS. By duality this system has a sequential limit $A^*$, which is dual to $A$. We now describe this RSS.

Again, for any $b\in S$, starting from $i$ sufficiently large, $b\in S_i\subseteq A^*_i$, and the sequence $\{||b||_i^*\}$ is monotone, however, in this case it is   non-decreasing. To show that it is convergent to a finite number we need to check that it is bounded from above. It is sufficient to check this for the basis vectors $\{e_i\}$, ${i\in I}$, since any sequence in $S_i$ is a finite sum of basis vectors. Let $\alpha\in I$, and $e_\alpha\in S$ be the corresponding basis vector. Then for any $i$ sufficiently large $||e_\alpha||^*_i=|\sup\limits_{||s||_i=1}|\langle e_\alpha,s\rangle |=|\langle e_\alpha,\frac{e_\alpha}{||e_\alpha||_i}\rangle|=\frac{1}{||e_\alpha||_i}\leq\frac{1}{||e_\alpha||}<\infty$,
since $||e_\alpha||=\inf\limits_i||e_\alpha||_i$ is by our assumptions greater than zero.

Thus we have a well-defined norm on $S$:
\be\label{norm_limit}
||a||^*_\infty=\lim\limits_{i\to\infty}||a||^*_i,
\ee
which gives rise to the RSS $A^*_\infty$. It turns out that  $A^*_\infty$ and $A$ are dual indeed. This is equivalent to the following.
\bt
The norms in (\ref{norm_colimit}) and (\ref{norm_limit}) are dual.
\et
{\bf Proof.}\hspace{0.01cm}

Let $a\in S$. We need to show that $||a||^*_\infty=\sup\limits_{b\in S, ||b||=1}|\langle a,b\rangle |$.

Let $V$ be the space of sequences with the same support as $a$, i.e. with zeros everywhere where $a$ is zero. Since $a$ is finitely supported the space $V$ is finite-dimensional. It is clear that all norms $||a||^*_i$, $|||a||^*_\infty$ are determined from pairing with elements of $V$.

Let $C=\{x\in V|\mbox{ }||x||=1\}$ be the unit sphere in $V$ for the norm $||.||$. Note that $C$ is compact.

Without loss of generality $||a||^*_\infty=1$. Pairing with $a$ is a continuous function on $C$, and by compactness of $C$ there is an $x\in C$ such that $|\langle a,x\rangle |$ is maximal on $C$, let $|\langle a,x\rangle |=K$. Pick $\epsilon>0$. Since $||x||=||a||^*_\infty=1$,  for all sufficiently large $i$ the norm $||x||_i<1+\epsilon$, and $||a||^*_i\leq1$, it follows that $\langle a,x\rangle |\leq ||a||^*_i||x||_i\leq||x||_i<1+\epsilon$. Since $\epsilon$ was arbitrary, it follows that $K\leq 1$.

On the other hand, again by compactness, for any $i$ there exists $z_i\in V$, $||z_i||_i=1$ such that $|\langle a,z_i\rangle |$ is maximal on the unit sphere for the norm $||\mbox{ }||_i$. That is, $|\langle a,z_i\rangle |=||a||^*_i$. Then $y_i=\frac{z_i}{||z_i||}\in C$ and $||y_i||_i=\frac{1}{||z_i||}$. We have  $|\langle a,y_i\rangle |=||a||^*_i\cdot||y_i||_i\geq ||a||^*_i\cdot||y_i||=||a||^*_i$. On the other hand, taking  $i$ sufficiently large, we may assume that $||a||^*_i>||a||-\epsilon=1-\epsilon$. So there exists $y\in C$  such that $|\langle a,y\rangle |\geq1-\epsilon$, and since $\epsilon$ was arbitrary if follows that $K\geq 1$.

So $\sup\limits_{y\in C}|\langle a,y\rangle |=K=1=||a||^*_\infty$. $\Box$
\bigskip

\bt In the setting as above the RSS $A$ is the sequential colimit of the system $\{A_i, f_{ij}\}$.
\et
{\bf Proof.}\hspace{0.01cm} The inclusion maps $A_i\to A$ obviously satisfy conditions of Lemma \ref{continuous-adjointable}, so they  are regular.  Since by definition $||a||\leq||a||_i$ for all $i$ and $a\in A_i$, the inclusion map $A_i\to A$ is a regular contraction, i.e. a morphism in ${\bf RSS}$, which fits in the commutative diagram with $f_{ki}$ for all $k\leq i$.

Let us check the universal property of colimit.

Assume $B$ is an RSS and we have a system of regular contracting maps $g_i:A_i\to B$, making together with the maps $f_{ij}$  a commutative diagram. We unambiguously define the map $g:S\to B$, putting $g(a)=g_i(a)$ for $a\in S$, where $i$ is sufficiently large, so that $a\in A_i$. By Note \ref{map_norm_finit} this map extends to a regular contracting map $g:A\to B$, since $\sup\limits_{||a||=1, a\in S} ||(|g|a)||=\sup\limits_{||a||\leq1, a\in S} ||(|g|a)||\leq\sup\limits_{||a||_i\leq1, a\in S_i} ||(|g|a)||=\sup\limits_i||g_i||\leq 1$. Obviously $g$ completes the commutative diagram,  and there is only one way to define such $g$. $\Box$
\bigskip

We have shown that in the described setting sequential colimit and, consequently, limit exist. Let us prove finally that the constructed colimit commutes with cotensor $\wp$  (hence, by duality the limit commutes with tensor $\otimes$). We will, however, consider commutation with  internal homs functor $\multimap$ rather than cotensor, which is the same thing anyway.

Consider the RSS $S_J\subseteq B\subseteq S_J^*$ on the index set $J$. This gives us the system of objects and maps $$\{B\multimap A_i, (id_B\multimap f_{ij}):B\multimap A_i\to B\multimap A_j\},$$ the maps being induced by the injections of the index sets $I_i\times J\subseteq I_j\times J$ for $i\leq J$. Hence this system has a sequential colimit, as we have just proven.

 There are two norms on the space of finite sequences $S_{I\times J}$: the norm $||.||$ of the internal hom-space $B\multimap A$ defined by (\ref{map_norm}), and the norm $||.||_\infty$ of the sequential colimit of the above system, defined by (\ref{norm_colimit}). It is sufficient to check that the two norms coincide.

Consider a finitely supported matrix $M\in S_{I\times J}$. We have
\be
||M||=\sup\limits_{a\in S_J,||a||=1}||(|M|a)||=\sup\limits_{a\in S_J,||a||=1}\inf\limits_i||(|M|a)||_i,
\ee
and
\be
||M||_\infty=\inf\limits_i\sup\limits_{a\in S_J,||a||=1}||(|M|a)||_i.
\ee
It follows that $||M||_\infty\geq||M||$ (since $\inf\limits_i\sup\limits_a\geq \sup\limits_a\inf\limits_i$).

Since we are interested only in the absolute value of $M$, assume without loss of generality that $M\geq 0$, so that $|M|=M$.

Let $V$ be the finite-dimensional space spanned by basis vectors $e_j\in S_J$  for which $Me_j\not=0$. Obviously both norms of $M$ are determined by their action on $V$.

Since $M$ is finitely supported, the image $MV$ is finite-dimensional, moreover for $i$ sufficiently  large $MV\subseteq S_i$. Hence, starting from $i$ sufficiently large, any norm $||.||_i$ is defined on $MV$, and, by finite-dimensionality,  all these norms are equivalent on this space.

Let $C\subset V$ be the unit sphere. Note that $C$ is compact. Let $||M||_\infty =K$

For every $i$ there exists  $x_i\in C$ such that $||Mx_i||_i=K_i$ is maximal on $C$, $K_i=||M||_i\geq K$. Let $y_i=Mx_i$. By compactness there is a converging subsequence of $\{x_i\}$; let $x\in C$ be its limit. By continuity of $M$, a subsequence of $\{y_i\}$ converges to $y=Mx$.

 Now pick $\epsilon>0$. There exists $N$, such that $||y||_i-||y||<\frac{\epsilon}{2}$ for all $i>N$. Also, since there is a subsequence of $\{y_i\}$, converging to $y$, and convergence in $MV$ is defined by any norm, in particular by the norm $||.||_N$, we have that there exists $i>N$, such that $||y_i-y||_N<\frac{\epsilon}{2}$. Then $||y_i-y||_i<\frac{\epsilon}{2}$ as well, since the sequence of norms is non-increasing. It follows that $||y||>||y||_i-\frac{\epsilon}{2}$ and $||y||_i>||y_i||_i-\frac{\epsilon}{2}$. So $||y||>||y_i||_i-\epsilon=K_i-\epsilon\geq K-\epsilon$. Since $\epsilon$ was arbitrary, we get $||Mx||=y\geq K=||M||_\infty$. It follows that $||M||\geq M_\infty$.

Thus $||M||=||M||_\infty$, and we have proven

\bt\label{colimits}
In the setting of this section the sequential colimit (respectively, limit) exists and commutes with cotensor (respectively, tensor)
\et

\subsection{Power series and free exponentials}
In the following we  use the notation
$$ \alpha(x_1\otimes\ldots\otimes x_k)=\alpha(x_1,\ldots, x_k)$$ and
$$x^{\otimes^i}=x\otimes\ldots\otimes x\mbox{ }i\mbox{ times}.$$

In the preceding sections we found the system of maps and objects $A_{\leq n}$, $i_{mn}:A_{\leq m}\to A_{\leq n}$ whose colimit $A_{\leq\infty}$,  if it exists and commutes with cotensor, is a free monoid in ${\bf RSS}$ by Theorem \ref{Mellies}. But, evidently, Theorem \ref{colimits} applies in this setting and, and we conclude that the free monoid $A_{\leq\infty}$ exists in ${\bf RSS}$.

 We are now going to give an explicit description of this space. It turns out, somewhat expectedly, that elements of $A_{\leq\infty}$ have interpretation as power series, absolutely converging in the unit ball of $A^*$, and we can identify $A_{\leq\infty}$ as $?A$.

In the following we denote the norm on $A_{\leq n}$ as $||.||_n$, and norm on $A_{\leq\infty}$ as $||.||$. From the preceding Section,  for the element $a\in A_{\leq m}$, its norm in $A_{\leq\infty}$ is
 \be ||a||=\lim\limits_{n\to\infty}||i_{mn}a||_n=\lim_{n\to\infty}\sup\limits_{i=1,\ldots,n, ||x_i||\leq 1}|(|i_{mn}a|(x_1,\ldots,x_n)|.
 \ee

 We show now that, in the limit $n\to\infty$, for the supremum in the above formula  we can assume all $x_1,\ldots,x_n$ equal.
More specifically, we show that, as $n\to\infty$, the value  $|i_{mn}a|(x_1,\ldots,x_n)$ almost coincides with $|i_{mn}a|((\frac{\sum x_i}{n})^{\otimes^n})$, the difference being bounded by an error estimate of order $O(\frac{1}{n})$ depending only on $a$ and $n$.

We are interested only in the absolute value $|a|$, so let us assume $a\geq 0$. Then $a=|a|$, $i_{mn}a\geq 0$,  and $|i_{mn}a|=i_{mn}a$.

The functional $a$ is expressible in terms of symmetric multilinear functionals $a_0,\ldots, a_m$ , see (\ref{multiaffine}). Let us estimate $|a_k((\frac{\sum x_i}{n})^{\otimes^k})|$.
\bl
For a symmetric $k$-linear functional $\alpha$ on $A^*$, and $x_i\in A^*$, $||x_i||\leq 1$, $i=1,\ldots, n$ it holds that $$\alpha((\frac{\sum x_i}{n})^{\otimes^k})=\frac{k!}{n^k}\sum\limits_{1\leq j_1<\ldots<j_k\leq n}\alpha(x_{j_1},\ldots,x_{j_k})+O(\frac{1}{n}),$$
as $n\to\infty$; the error term $O(\frac{1}{n})$ depending only on $\alpha$ and $n$.
\el
{\bf Proof.}\hspace{0.01cm}
From multinomial expansion we have: $$\alpha((\frac{\sum x_i}{n})^{\otimes^k})=
\frac{1}{n^k}\sum\limits_{\substack{i_1,\ldots,i_n\geq 0,\\ i_1+\ldots+i_n=k}}\frac{k!}{i_1!\ldots,i_n!}\alpha(x_1^{\otimes^{i_1}}\otimes\ldots\otimes x_n^{\otimes^{i_n}})=$$
$$=\frac{1}{n^k}\sum\limits_{r=1}^k\sum\limits_{1\leq j_1<\ldots<j_r\leq n}\sum\limits_{\substack{i_1,\ldots,i_r\geq1,\\ i_1+\ldots+i_r=k}}\frac{k!}{i_1!\ldots, i_r!}\alpha(x_{j_1}^{\otimes^{i_1}}\otimes\ldots\otimes x_{j_r}^{\otimes^{i_r}})=$$ $$=\frac{1}{n^k}\sum\limits_{r=1}^{k-1}(\cdots)+\frac{k!}{n^k}\sum\limits_{1\leq j_1<\ldots<j_k\leq n}\alpha(x_{j_1},\ldots,x_{j_k}).$$

Let us estimate the first sum in the RHS of the above formula:
$$|\frac{1}{n^k}\sum\limits_{r=1}^{k-1}\sum\limits_{1\leq j_1<\ldots<j_r\leq n}\sum\limits_{\substack {i_1,\ldots,i_r\geq1,\\ i_1+\ldots+i_r=k}}\frac{k!}{i_1!\ldots i_r!}\alpha(x_{j_1}^{\otimes^{i_1}}\otimes\ldots\otimes x_{j_r}^{\otimes^{i_r}})|\leq $$
$$\leq||\alpha||\frac{1}{n^k}\sum\limits_{r=1}^{k-1}\sum\limits_{1\leq j_1<\ldots<j_r\leq n}\sum\limits_{\substack{i_1,\ldots,i_r\geq1, \\i_1+\ldots+i_r=k}}\frac{k!}{i_1!\ldots i_r!}=$$ $$=||\alpha||\frac{1}{n^k}\sum\limits_{r=1}^{k-1}\frac{n!}{r!(n-r)!}r^k\leq
||\alpha||\frac{1}{n^k}(\sum\limits_{r=1}^{k-1}\frac{n!}{r!(n-r)!})(\sum\limits_{r=1}^{k-1}r^k)=$$
$$=
||\alpha||\frac{1}{n^k}\cdot\{\mbox{degree  }k-1\mbox{ polynomial in }n\}\cdot\{\mbox{ constant independent of }n\}=O(\frac{1}{n}).$$
The bound $O(\frac{1}{n})$ depends only on $n$ and $\alpha$, and the statement follows. $\Box$
\bigskip

From (\ref{multiaffine}) and (\ref{i_mn}), we have for $a\in A_{\leq m}$, $a\geq 0$:
$$(i_{mn}a)(x^{\otimes^n})=\sum\limits_{k=0}^{m}\frac{n!}{k!(n-k)!}a_k(x^{\otimes^k})\frac{m!(n-k)!}{n!(m-k)!}=
\sum\limits_{k=0}^{m}\frac{m!}{k!(m-k)!}a_k(x^{\otimes^k}).$$

Then, using the preceding Lemma, we get: $$(i_{mn}a)((\frac{\sum x_i}{n})^{\otimes^n})=\sum\limits_{k=0}^{m}(\frac{m!}{n^k(m-k)!}\sum\limits_{1\leq i_1<\ldots<i_k\leq n}a_k(x_{i_1},\ldots,x_{i_k})+O(\frac{1}{n}))=$$ $$=\sum\limits_{k=0}^{m}\frac{m!}{n^k(m-k)!}\sum\limits_{1\leq i_1<\ldots<i_k\leq n}a_k(x_{i_1},\ldots,x_{i_k})+O(\frac{1}{n}),$$ the term $O(\frac{1}{n})$ depending only on $n$ and norms $||a_k||$, $k=1,\ldots, m$.

On the other hand,   $$\frac{(n-k)!}{n!}=\frac{1}{n^k}+O(\frac{1}{n^{k+1}})\mbox{ as }n\to\infty,$$ and it follows from (\ref{i_mn}) and the preceding estimates that
$$
(i_{mn}a)(x_1,\ldots,x_n)=\sum\limits_{k=0}^{m}(\frac{m!}{n^k(m-k)!}\sum\limits_{1\leq i_1<\ldots<i_k\leq n}a_k(x_{i_1},\ldots,x_{i_k}))+$$
$$+\sum\limits_{k=0}^{m}(O(\frac{1}{n^{k+1}})\frac{m!}{(m-k)!}\sum\limits_{1\leq i_1<\ldots<i_k\leq n}a_k(x_{i_1},\ldots,x_{i_k})).$$
The error term in the righthand side of the above can be estimated as less than or equal to $\sum\limits_{k=0}^{m}O(\frac{1}{n^{k+1}})\frac{m!}{(m-k)!}\frac{n!}{k!(n-k)!}=O(\frac{1}{n})$, and we conclude that
$$
(i_{mn}a)(x_1,\ldots,x_n)=(i_{mn}a)((\frac{\sum x_i}{n})^{\otimes^n})+O(\frac{1}{n}).
$$

This means that the norm $||a||$ of $a$ in $A_{\leq\infty}$ is determined by values of $a$ on the unit ball of $A^*$, when $a$ is seen as the polynomial function of one variable
\be
a(x)=(n!)\sum\limits_{k=0}^n\frac{a_k(x^{\otimes^k})}{k!(n-k)!},
\ee
with the usual abuse of notation.
\bt
For $a\in A_{\leq m}$, its norm in $A_{\leq\infty}$ is defined by (\ref{power_series_norm}).
$\Box$
\et
\bigskip

It follows that $A_{\leq \infty}$ is the RSS on the index set $M(I)$ defined by norm (\ref{power_series_norm}), and this is identified as $?A$. But by Theorem \ref{Mellies}, the space $A_{\leq\infty}$ is a free monoid over $A$ and we conclude with the following.
\bt
The RSS $?A$ of power series, absolutely converging on the unit ball of $A^{*}$ is a free monoid over $A$. $\Box$
\et
\bigskip

By Theorem \ref{Mellies} this yields us the following
\bc
The RSS $!A$, $?A$ give a model of linear logic exponential connectives. $\Box$
\ec

\section{Some remarks and questions}
We thus get a vector spaces model of linear logic with free exponentials. Its drawbacks, however, are apparent.

Our spaces are  explicitly defined as sequence spaces, i.e. with the use of a particular coordinate system. We tried to be as coordinate-free as possible; that is one of the reasons why we often represent an RSS as an abstract ``sandwich'' $S\subseteq A\subseteq S^*$, rather than a more concrete space of sequences.  Unfortunately, constructions of the current work still crucially rely on index sets, and it remains a question if we can do something more interesting.

In fact, the lattice structure of sequence spaces is also heavily used in our work. In principle, it seems possible to give a coordinate-free presentation, describing our sequences spaces as very specific vector lattices. Such a possibility would be interesting, however, if we were able to generalize to some wider class of vector lattices or, more generally, partially ordered vector spaces. This is a topic of further research.
For example, it would  be exciting to consider {\it ``non-commutative sequences''} spaces, i.e. spaces of self-adjoint operators.



\begin{thebibliography}{17}
\bibitem{BluteEhrhardTasson} R. Blute, T. Ehrhard, C. Tasson, A Convenient Differential Category, Cahiers de Topologie et Geometrie Differentielle 53, pp. 211-232, 2012.

\bibitem{BlutePanangadedOldFoundations} R. Blute, P. Panangaden and R. Seely, Old Foundations for Linear Logic, Proceedings
of the 9th Symposium on Mathematical Foundations of Programming Semantics, April
1993.  LNCS 802, pp.  474-512.
\bibitem{Seely2} J.R.B. Cockett, R.A.G. Seely, Proof theory for full intuitionistic linear logic, bilinear logic, and mix categories, Theory and Application of Categories 3(5) (1997) 85-131.
\bibitem{DanosErhahrdPCS} V. Danos, T. Ehrhard, Probabilistic coherence spaces as a model of higher-order probabilistic computation, Information and Computation 209, Issue 6, 966-991, 2011
\bibitem{Ehrhard_Koethe} Th. Ehrhard, On Koethe sequence spaces and  linear         logic.
Mathematical         Structures         in
Computer        Science,        12:579-623,        2002.
\bibitem{diff-lambda}   T. Ehrhard, L. Regnier, The differential lambda-calculus. Theoretical Computer Science 309(2003) 1-41.
\bibitem{diff-nets} T. Ehrhard, L. Regnier, Differential interaction nets. Theoretical
Computer Science 364(2006) 166-195.
\bibitem{GelfandVilenkin} I. M. Gelfand and N. J. Vilenkin. Generalized Functions, vol. 4: Some Applications of Harmonic Analysis. Rigged Hilbert Spaces. Academic Press, New York, 1964.
\bibitem{Girard} J.-Y. Girard, Linear logic, Theoretical Computer Science 50 (1987) 1-102.
\bibitem{Girard2} J.-Y. Girard, Linear logic: its syntax and semantics, in J.-Y.Girard, Y.Lafont and L.Regnier, eds. Advances in Linear Logic, 1-42, Cambridge University Press, 1995, Proc. of the Workshop on Linear Logic, Ithaca, New York, June, 1993.
\bibitem{Girard99coherentbanach}
     {J.-Y. Girard}, Coherent Banach spaces: a continuous denotational semantics,
     {Theoretical Computer Science},
    227,
    275-297, (1999).
\bibitem{Girard_Logic_quantic} J.-Y.  Girard. Between logic and quantic: a tract.  In            Thomas            Ehrhard,            Jean-Yves
Girard,       Paul       Ruet,       and       Philip       Scott,       editors,
Linear        Logic        in        Computer        Science
,       volume       316
of
London         Mathematical         Society         Lecture         Note         Series
,        pages        346-381.        Cambridge        University
Press,        2004.
\bibitem{Lafont} Y. Lafont,  Logique, cat\'egories et machines, Th\`ese de doctorat, Universit\'e de Paris 7,
Denis Diderot,  1988.
\bibitem{Mellies_free_exp} P.-A. Melli\'es, N. Tabareau, C. Tasson, An explicit formula for the free exponential
modality of linear logic, Lecture Notes in Computer Science,  5556,  247-260, 2009.
\bibitem{Mellies_categorical_semantics} P.-A. Melli\'es, Categorical semantics of linear logic, in: Interactive Models of Computation and Program Behaviour, Panoramas et Synth\`eses 27, Soci\'et\'e Math\'ematique de France 1–196, 2009.
\bibitem{Rudin} {Rudin, Walter},{Functional analysis}, {International Series in Pure and Applied Mathematics},
  {McGraw-Hill Inc.}, {New York}, {1991}.
\bibitem{Seely}  R.A.G. Seely, Linear logic, $*$-autonomous categories and cofree coalgebras, in: J.Gray and A.Scedrov (editors), Categories in Computer Science and Logic, Contemporary Mathematics 92, 371-382, Amer. Math. Soc., 1989.

\end{thebibliography}
 \end{document}